\documentclass[pss]{wiley2sp} 
\usepackage{amsmath}

\begin{document}

\title{Carrier Multiplication in Silicon Nanocrystals:  Theoretical  Tools and Role of the Passivation}

\titlerunning{Short title }

\author{%
  Ivan Marri\textsuperscript{\Ast,\textsf{\bfseries 1}},
  Marco Govoni \textsuperscript{\textsf{\bfseries 2}},
  Stefano Ossicini \textsuperscript{\textsf{\bfseries 3}}}

\authorrunning{First author et al.}

\mail{e-mail
  \textsf{marri@unimore.it}, Phone:
  +39 205 5067, Fax: +39 059 2055651}

\institute{%
  \textsuperscript{1}\,Centro S3, CNR-Istituto Nanoscienze, via Campi 213/a
Modena, 41125, Italy
\\
  \textsuperscript{2}\, Institute for Molecular Engineering and Materials Science Division, Argonne National Lab 9700 Cass Avenue, Argonne,  USA
\\
  \textsuperscript{3}\,University of Modena and Reggio Emilia, Department of Science and Methods for Engineering (DISMI), via Amendola 2 Reggio Emilia, 42122, Italy}



\abstract{%
%
%
%
\abstcol{%
 Carrier multiplication is a non-radiative recombination mechanism that leads to the generation of two or more electron-hole pairs after absorption of a single photon. By reducing the occurrence of dissipative effects, this process can be exploited to increase solar cell performance. In this work we introduce two different theoretical  fully ab-initio tools that can be adopted to study carrier multiplication in nanocrystals. The tools are described in detail and compared.  Subsequently we calculate carrier multiplication  lifetimes in H- and OH- terminated silicon nanocrystals, pointed out the role played by the passivation on the carrier multiplication processes.
}}

%
%
\titlefigure[width=\linewidth, scale=0.6 ]{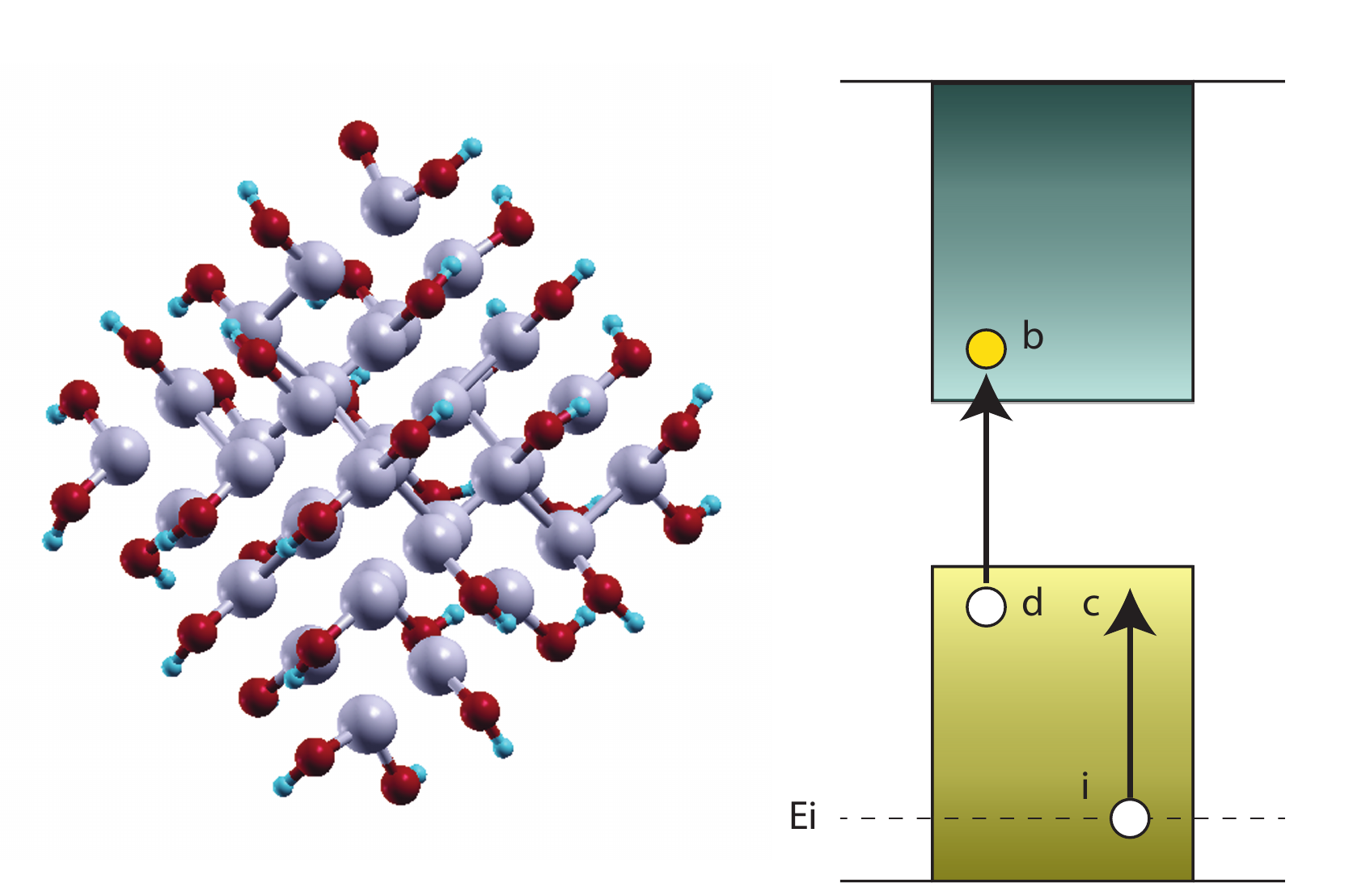}
\titlefigurecaption{%
 The $\text{Si}_{35}\text{(OH)}_{36}$ is depicted in the left part of the figure. On the right we schematize the a Carrier Multiplication decay process ignited by the relaxation of a hole.}

\maketitle   

\section{Introduction}
The advent of nanoscience and nanotechnologies has opened new prospectives in the development of novel nanomaterials. 
Control and manipulation of matter on an atomic scale can be exploited to generate novel nanostructured systems
where new properties and functionalities are promoted to overcome the limits of traditional semiconductor and organic systems for applications in different fields, from medical devices to drags, from photonics to energy conversion \cite{marri_SSC,GUERRA_SM,Ossicini_JNN,DEGOLI2_CRP,iori_NC1,Iacomino_PRB,iori_doping,guerra}.
At the nanoscale, electronic, optical and transport properties can be tuned by size reduction, by changing passivation and shape of the nanostructures, by doping or by surface functionalization. 
In particular, effects induced by the size reduction can be exploited to promote new  recombination processes that are negligible in bulk-like systems. This is the case of the non-radiative Coulomb-driven Carrier Multiplication (CM), that is the counterpart of the Auger recombination \cite{govoni_augerbulk}.
This effect leads to the generation of  multiple electron-hole (e-h) pairs  after absorption of a single high energy photon, with an energy at least twice the energy gap of the system. 
By reducing the occurrence of high-energy  dissipative processes and by increasing the number of carrier generated after photon absorption,  CM is expected to increase solar cell efficiency.
CM has been recorded in a large variety of NCs like for instance PbSe and PbS\cite{ellingson_exp_PbSe_PbS,schaller_exp_PbSe,trinh_exp_PbSe,nair_exp_PbSe_PbS,schaller_seven,schaller_exp_PbSe_CdSe}, CdSe and CdT \cite{schaller_exp_PbSe_CdSe,schaller_exp_CdSe,gachet_exp_coreshell}, PbTe \cite{murphy_exp_PbTe}, InAs \cite{schaller_exp_InAs} , Si \cite{beard_exp_Si} and  Ge \cite{Saeed}.  Recently a relevant photocurrent enhancement arising from CM was proven in a PbSe based quantum dot solar-cell \cite{Semonin_CM}.  Effects induced on CM dynamics by energy transfer quantum cutting processes were observed by Timmerman et al. \cite{timmerman,timmerman_nnano} and Trinh et al. \cite{trinh} in Si-NCs organized in a dense array, despite the presence of the observed effect is still under discussion \cite{valenta_SSQC}.
On a theoretical side, CM have been often described as an  impact ionization (II) process that follow the primary photoexcitation event.  This approach was combined with a semi-empirical tight binding model  by  G. Allan and C. Delerue \cite{delerue_impact1,delerue_impact2,allan_stativuoto,delerue_CMimpact} to calculate CM rates in PbSe, PbS and Si nanocrystals (NCs) and with a semiempirical pseudopotential method by E. Rabani and R. Bauer to estimate biexciton generation rates in  CdSe and InAS NCs \cite{rabani_nanolett_CdSe_InAs}. Similar calculations were also performed by M. Califano et al. (semi-empirical nonlocal pseudopotential approach) \cite {califano_CdSe,califano_apl} and by A. Franceschetti et al.  \cite{franceschetti_CMimpact,Franceschetti_PRL} (atomistic pseudopotential method) to obtain CM rates in CdSe and PbSe NCs. Recently a new full ab-initio approach was adopted by M. Govoni et al. \cite{govoni_nat} and by I. Marri et al. \cite{marri_JACS,Marri_Beilstein,MARRI_SOLMAT} to calculate CM lifetimes in systems of isolated and strongy interacting Si-NCs.\\
\noindent
In this work, we present new fully ab-initio calculations of CM lifetimes for both spherical hydrogenated and oxygenated Si-NCs.  We discuss  results obtained using two different theoretical approaches and we investigate, for the first time, CM dynamics in  OH-terminated Si-NCs.
\section{Method}
\label{Method}
\begin{figure}[t]%
\includegraphics*[width=\linewidth]{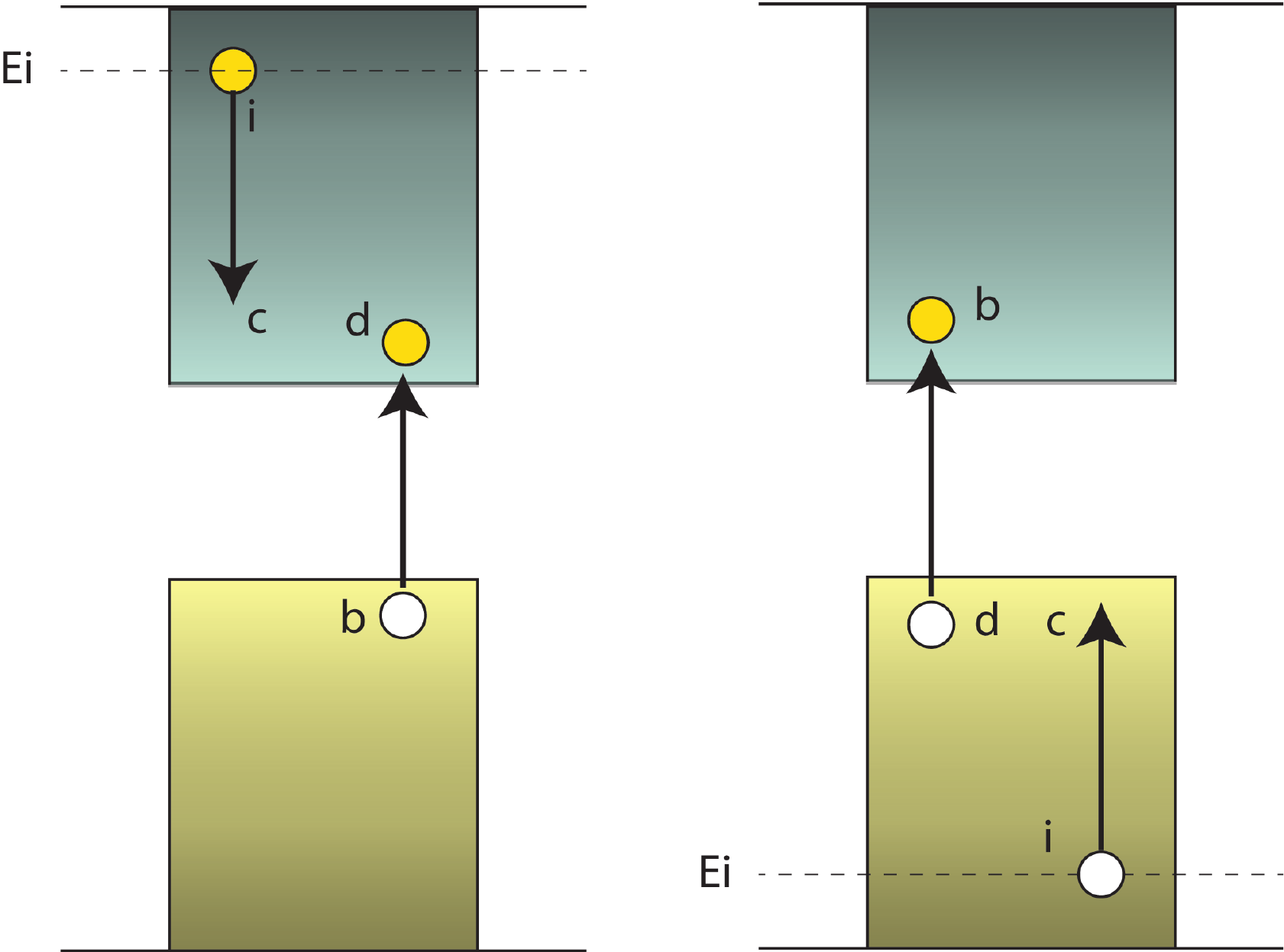}
\caption{%
CM processes ignited by electron (left) and hole (right) relaxation are depicted in the figure.}
\label{Fig_1}
\end{figure}

In this section we describe the methodologies adopted to calculate CM lifetimes.  Following  Refs. \cite{delerue_impact1,rabani_nanolett_CdSe_InAs,allan_stativuoto}, recombination by  CM decays is described as an II process that occurs after  the absorption of a single high-energy photon ($h\nu > 2E_{g}$, where $E_{g}$ is the energy gap of the system). In this scheme, the decay of an high energy exciton into a biexciton is  modeled as the sum of two processes, one ignited by electron relaxation that decays in a negative trion (the hole is a spectator, see Fig. \ref{Fig_1} left decay scheme),  and one ignited by hole relaxation that decays in a positive trion (the electron is a spectator, see Fig.  \ref{Fig_1} right decay scheme).
In this work, electronic structures are calculated from first-principle using the Density Functional Theory DFT. In particular, Kohn-Sham (KS) states are obtained by solving the KS equation:
\begin{equation}
\hat{H}^{\sigma}_{KS} \psi_{n k \sigma}=\epsilon_{n k \sigma} \psi_{n k \sigma} 
\label{KS}
\end{equation}
where  the  hamiltonian $\hat{H}^{\sigma}_{KS}$  is given by the sum of four terms, the kinetic energy operator and the ionic, Hartree and exchange-correlation potentials.
\noindent
The CM rate $R_{n_a, {\bf k}_a}^{e}(E_i)$ for mechanisms ignited by an electron relaxation are given by:
\begin{eqnarray}
&& R_{n_a, {\bf k}_a}^{e}(E_i)= \sum_{n_c,n_d}^{cond.} \sum_{n_b}^{val.} \sum_{{\bf{k}}_b, {\bf{k}}_c, {\bf{k}}_d}^{1BZ}  4 \pi \Big[ \mid M_D \mid^{2} + \mid M_E \mid^{2} \nonumber \\
&&  + \mid M_D - M_E \mid^{2} \Big] \delta(E_a+E_b-E_c-E_d).
\label{CMe}            
\end{eqnarray}
Similarly, for mechanisms induced by hole relaxation, we have:
\begin{eqnarray}
&& R_{n_a, {\bf{k}}_a}^{h}(E_i)= \sum_{n_c,n_d}^{val.} \sum_{n_b}^{cond.} \sum_{{\bf{k}}_b, {\bf{k}}_c, {\bf{k}}_d}^{1BZ}  4 \pi \Big[ \mid M_D \mid^{2} + \mid M_E \mid^{2} \nonumber \\
&&  + \mid M_D - M_E \mid^{2} \Big] \delta(E_a+E_b-E_c-E_d) 
\label{CMh}            
\end{eqnarray} 
where the indexes n and k identify KS states, $E_{i}$ is the energy of the carrier that ignites the decay CM mechanism, 1BZ is the first Brillouin zone and $\mid M_{D} \mid$ and $\mid M_{E} \mid$ are the direct and exchange screened Coulomb matrix elements, respectively. In our simulations, the delta function for energy conservation was implemented in the form of a Gaussian distribution with a full width at half maximum of $0.02$ eV.
 In reciprocal space, $\mid M_{D} \mid$ and $\mid M_{E} \mid$ assume the form:
$$
\text{M}_D= \frac{1}{V} \sum_{{\bf{G}}, {\bf{G}}'} \rho_{n_d, n_b}({\bf{k}}_d, {\bf{q}}, {\bf{G}}) \text{W}_{{\bf{G}} {\bf{G}}'}  \rho_{n_a, n_c}^{\star}({\bf{k}}_a, {\bf{q}}, {\bf{G}'}) \
$$
$$
\text{M}_E=
\frac{1}{V} \sum_{{\bf{G}}, {\bf{G}}'} \rho_{n_c, n_b}({\bf{k}}_c, {\bf{q}}, {\bf{G}}) \text{W}_{{\bf{G}} {\bf{G}}'}  \rho_{n_a, n_d}^{\star}({\bf{k}}_a, {\bf{q}}, {\bf{G}'}) \
$$
where both ${\bf{k}}_c + {\bf{k}}_d -{\bf{k}}_a -{\bf{k}}_b$ and ${\bf{G}}, {\bf{G}}'$ are vectors of the reciprocal space, $\bf{q}=({\bf{k}}_c -{\bf{k}}_a)_{1BZ}$ and 
$\rho_{n, m}({\bf{k}}, \bf{q}, {\bf{G}})= \langle n, {\bf{k}} \vert e^{i( {\bf{q}}+ {\bf{G}})\cdot {\bf{r}}} \vert m,  {\bf{k}} - {\bf{q}} \rangle$ is the oscillator strength.
The Fourier transform of the screened interaction $W_{{\bf{G}}, {\bf{G}}'}({\bf{q}}, \omega)$, identified by a matrix in ${\bf{G}}$ and ${\bf{G}}'$ is evaluated at $\omega=0$ and assume the form;
\begin{eqnarray}
&&W_{{\bf{G}}, {\bf{G}}'}({\bf{q}}, 0) = v_{{\bf{G}}, {\bf{G}}'}^{bare}({\bf{q}})+W^{p}_{{\bf{G}}, {\bf{G}}'}({\bf{q}}, 0) =\nonumber \\
&& \frac{4 \pi \cdot \delta_{{\bf{G}}, {\bf{G}}'}}{\mid {\bf{q}}+ {\bf{G}} \mid^2} +
\frac{\sqrt{4 \pi e^2}}{\mid {\bf{q}}+ {\bf{G}} \mid} \bar{\chi}_{{\bf{G}}, {\bf{G}}'}({\bf{q}}, 0) \frac{\sqrt{4 \pi e^2}}{\mid {\bf{q}}+ {\bf{G}'} \mid} \nonumber \\
&&
\label{W}            
\end{eqnarray}
The first term ($v_{{\bf{G}}, {\bf{G}}'}^{bare}({\bf{q}})$) of Eq. \eqref{W} denotes the bare interaction while the second one ($W^{p}_{{\bf{G}}, {\bf{G}}'}({\bf{q}}, \omega)$) includes the screening caused by the medium. 
In Eq. \eqref{W}, $\bar{\chi}_{{\bf{G}}, {\bf{G}}'}({\bf{q}}, \omega)$ is the symmetrized reducible polarizability. Noticeably,  $W^{p}_{{\bf{G}}, {\bf{G}}'}({\bf{q}}, \omega)$ is often given as a function of the  reducible polarization $\chi_{{\bf{G}}, {\bf{G}}'}({\bf{q}}, \omega)$, where: 
$$
\bar{\chi}_{{\bf{G}}, {\bf{G}}'}({\bf{q}}, 0)=\frac{\sqrt{4 \pi e^2}}{\mid {\bf{q}}+ {\bf{G}} \mid}  \chi_{{\bf{G}}, {\bf{G}}'}({\bf{q}}, \omega=0) \frac{\sqrt{4 \pi e^2}}{\mid {\bf{q}}+{\bf{G}}' \mid} 
$$
The reducible polarizability is connected to the irreducible polarizability ${\chi}_{{\bf{G}}, {\bf{G}}'}^0$ by the Dyson equation that, in the Random Phase Approximation (RPA), assumes the form:
$$
{\chi}_{{\bf{G}}, {\bf{G}}'}={\chi}_{{\bf{G}}, {\bf{G}}'}^0 +\sum_{G_1, G_2} {\chi}_{{\bf{G}}, {\bf{G}_1}}^0 v_{{\bf{G}_1}, {\bf{G}_2}}{\chi}_{{\bf{G}_2}, {\bf{G}}'}
$$ 
The presence of off-diagonal elements in the solution of the Dyson equation is related to the inclusion of the local fields (LFs) that stem from the breakdown of the translational invariance imposed by the lattice.
Different strategies have been adopted to calculate Eq. \eqref{W}, and in particular the term $W^{p}_{{\bf{G}}, {\bf{G}}'}({\bf{q}}, \omega)$. Two of them will be here discussed  and then applied to calculate CM lifetimes in Si-NCs.\\
\noindent
The first one permits to develop the noninteracting response function ${\chi}_{{\bf{G}}, {\bf{G}}'}^0$ in terms of the bare Green's function, that is:
\begin{eqnarray}
&&{\chi}_{{\bf{G}}, {\bf{G}}'}^0({\bf{q}}, \omega)=2\sum_{n,n'} \int_{BZ}\rho_{n', n}^{\star}({\bf{k}}, {\bf{q}}, {\bf{G}})\rho_{n', n}({\bf{k}}, {\bf{q}}, {\bf{G}'}) \times\nonumber \\
&& \left[  \frac{f_{n {\bf{k-q}}}(1-f_{n' {\bf{k}}})}{\omega+\epsilon_{n, {\bf{k}}-{\bf{q}} }  - \epsilon_{n',{\bf{k}} } +i0^+}  -  \frac{f_{n {\bf{k-q}}}(1-f_{n' {\bf{k}}})}{\omega+ \epsilon_{n',{\bf{k}} } -\epsilon_{n, {\bf{k}}-{\bf{q}} }  - i0^+} \right] \nonumber \\
&&
\label{chi}
\end{eqnarray}
where $f_{n {\bf{k}}}$ is the occupation factor of the $\mid n  {\bf{k}}>$ state.
This method will be refereed as the bare Green's function  (BGF) procedure in the following.
We applied this procedure in the  works of Refs. \cite{govoni_nat,marri_JACS}, where a modified version of the code Yambo \cite{Marini} was used to calculate CM rates in systems of isolated and interacting Si-NCs.
Relation (\ref{chi}) implies a sum over both occupied and unoccupied states. Due to the complexity of the procedure, that leads to the calculation of a large number of matrix elements (this number depends on the size of the system, on the number of electrons, on the kinetic energy cutoff used for the wavefunctions  etc..), methodology of Eq. \eqref{chi} often requires a truncation  of the reducible density-density $\chi_{{\bf{G}}, {\bf{G}}'}$. In Refs. \cite{govoni_nat}, for instance, we imposed  energy  cutoffs ranging from 1.5 to 0.5 Hartree and we obtained, for the largest systems  $\chi_{{\bf{G}}, {\bf{G}}'}$ matrices with a size of about $22500 \times 22500$.
\\
\noindent
The second one,  defined projective eigendecomposition of the dielectric screening (PDEP) \cite{Nguyen_PRB,Pham_PRB,Govoni_JCTC,Wilson_PRB_1,Wilson_PRB_2}, implies a spectral decomposition of the symmetrized irreducible polarizability,  that occurs through an iterative projection of the eigenvectors of $\bar{\chi}_{{\bf{G}}, {\bf{G}}'}^0({\bf{q}},0)$ via repeated linear-response computations within the density functional perturbation theory (DFPT) and iterative diagonalization algorith \cite{Baroni_PRL,Baroni_RMP}. In this approach the symmetrized irreducible polarizability is developed as:
\begin{equation}
\bar{\chi}_{{\bf{G}}, {\bf{G}}'}^0({\bf{q}, 0})= \sum_{i,j}^{N_{pdep}} \phi_{i}({\bf{q}}+{\bf{G}}) \lambda_{i,j}\phi_{j}^{\star}({\bf{q}}+{\bf{G}'})
\label{PDEP}
\end{equation}
where the $N_{pdep}$ $\phi_{i}$ vectors are iteratively calculated by solving the Sternheimer equation \cite{Sternheimer} without explicitly evaluating empty states. Initially 
 a set of orthogonal $N_{pdep}$ basis functions with random components is defined. Then a given perturbation $V_i^{pert}=\phi_{i}$ is applied  and the 
 linear variation $\Delta \psi_{n k \sigma}$ of the  occupied eigenstates $\psi_{n k \sigma}$ is calculated using the relation:
\begin{equation}
(H_{KS}^{\sigma} - \epsilon_{n {\bf{k}} \sigma}) \hat{P}_{c}^{{\bf{k}} \sigma} \Delta \psi_{n k \sigma}=-\hat{P}_{c}^{{\bf{k}} \sigma} V_i^{pert}
 \psi_{n k \sigma}
 \label{baroni}
\end{equation}
where $\hat{P}_{c}^{{\bf{k}} \sigma}$ is  the projector operator over the occupied manyfold of state with momentum ${\bf{k}}$ and spin  $\sigma$.
Finally   Eq. (\ref{baroni}) is solved iteratively by using for instance the preconditioned conjugate-gradient method. The linear variation of the density induced by the i-th perturbation is then obtained by:
\begin{equation}
\Delta n_{i}(r)=\sum_{\sigma}\sum_{n=1}^{N_{occ}^{\sigma}} \int_{BZ} \frac{d {\bf{k}}}{(2 \pi)^3} \left[ \psi_{n k \sigma}(r)^{\star} \Delta \psi_{n k \sigma}(r)^{i} +cc \right]
\end{equation}
and  the matrix elements of the irreducible polarizability in the space spanned by $\phi_{i}$ are calculated using the relation $\bar{\chi}_{ij}^{0}= < \phi_{i} \mid  \Delta n_{j}(r) > $. The matrix is then diagonalized to obtain a new set of basis vectors $\phi_i$ and the procedure is iterated using a Davidson algorithm.\\
\noindent
After the calculation of both direct and exchange Coulomb matrix elements using  both the BGF and PDEP procedures, CM rates are obtained (Eq. \eqref{CMe} and \eqref{CMh}). CM lifetimes are then calculated as reciprocal of rates  and are given as a function of the energy of the initial carrier $E_i$. \\
\section{Results}
We analyze  hydrogenated and the oxygenated Si-NCs. These systems are obtained by terminating the surface dangling bonds with H- and with OH- (hydroxyls functional group), respectively. 
This section is divided in two parts. Initially we apply and compare the methods described in Section \ref{Method} to calculate CM lifetimes for a simple H-terminated Si-NCs. Subsequently we extend our analysis to consider an oxygen passivated Si-NCs;  CM lifetimes are then calculated using the PDEP procedure. To our knowledge, it is the first time that a fully ab-initio calculation is performed to evaluate effects induced by the passivation on the CM activity.
We consider the $\text{Si}_{35}\text{H}_{36}$ \cite{govoni_nat} and the $\text{Si}_{35}\text{(OH)}_{36}$.  As pointed out in Ref. \cite{guerra}, this configuration well describe a system formed by a Si-NC encapsulated in a $\text{SiO}_2$ matrix. 
The eletronic structure is calculated using the density functional theory plane-wave pseudopotential PWscf code of the Quantum-ESPRESSO package \cite{giannozzi}; silicon, hydrogen and oxygen norm-conserving pseudopotentials are adopted and the local density approximation LDA is used to  calculate the exchange-correlation functional.  
For the fully hydrogenated system we adopt  a kinetic energy cutoff  for wavefunctions of $35$ Ry; this value move to $55$ Ry when the OH terminated system is considered. A  simple cubic cell with a lattice parameter of 90 atomic units (a.u.) is adopted for the H-terminated system (this values is reduced to 70 a.u. when the PDEP procedure is adopted, in order to reduce the number of G vectors and therefore the computational cost of the simulation), while a cubic FCC cell with a lattice parameters of $95$ a.u.  is employed for the OH-terminated NC.
\subsection{H-terminated Si-NCs: a direct comparison between two different tools}
\label{H_term}
\begin{figure}[t]%
\includegraphics*[width=\linewidth]{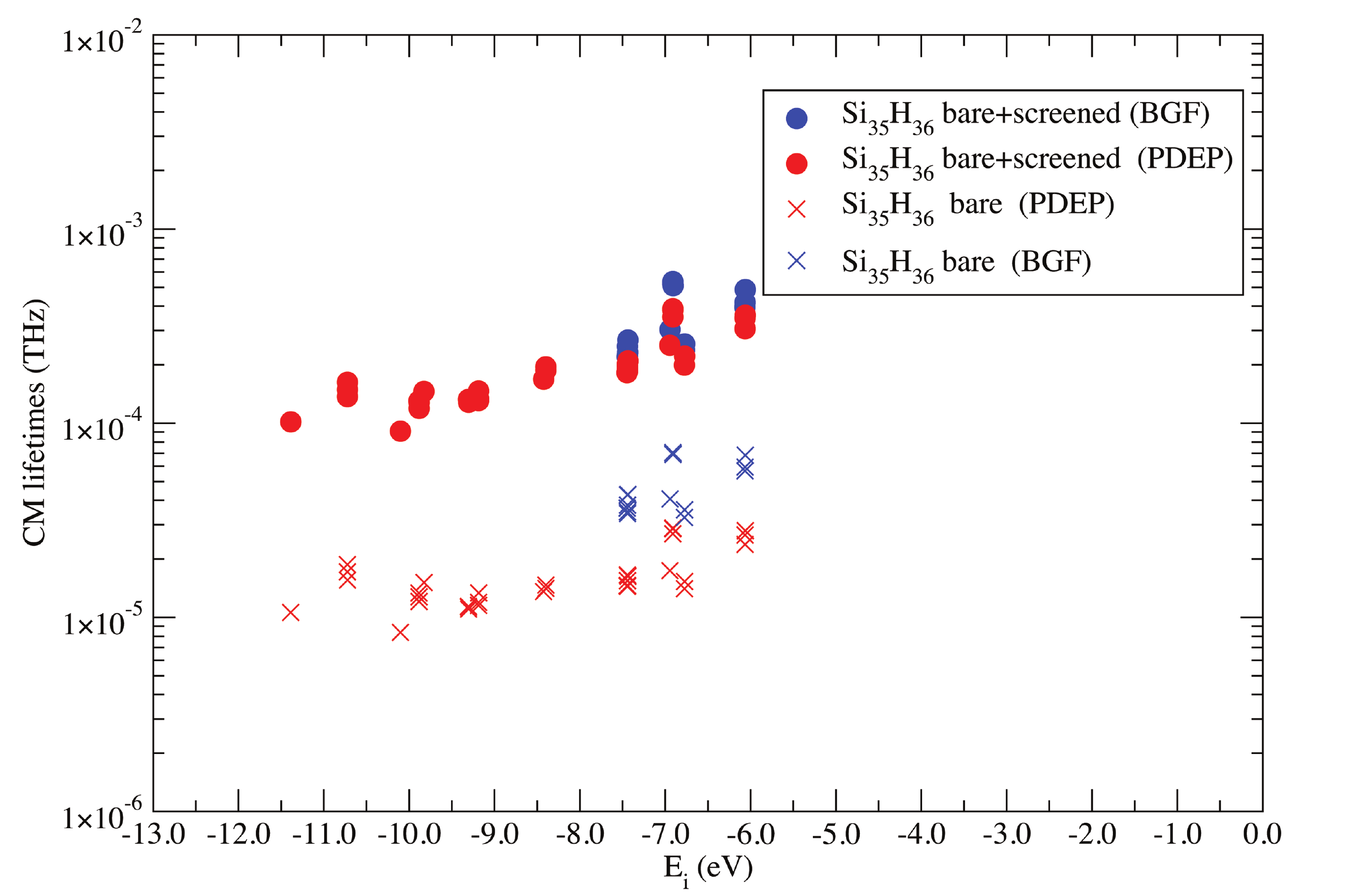}
\caption{%
CM processes ignited by electron (left) and hole (right) relaxation are depicted in the figure.}
\label{Fig_2}
\end{figure}

In this section we report the calculated CM lifetimes for the  $\text{Si}_{35}\text{H}_{36}$ NC. In particular, we compare the  results obtained with  procedures described in section \ref{Method} , that is the so called BGF method  (Eq. \eqref{chi}) (already adopted by us  in the works of  Ref. \cite{govoni_nat}) and the PDEP procedure, Eq \eqref{PDEP}.  The methodology of Eq. \eqref{chi} have been applied by imposing an energy cutoff of $1.5$ Hartree to the screening term. A spectral decomposition on $1300$ basis vectors is instead adopted in the PDEP methodology. \\
\noindent
Calculated CM lifetimes  for mechanisms ignited by hole relaxation are reported in the Fig. \ref{Fig_2}. Here different colors indicated the different methodology used in the simulation. Crosses and dots instead identify the lifetimes calculated considering only the bare term of Eq.  \eqref{W} or both the bare and the screened term of Eq.  \eqref{W}.
Following the conclusions of Ref. \cite{govoni_nat} we do not consider contributions due to the vacuum states. As a consequence for the $\text{Si}_{35}\text{H}_{36}$ electron-initiated CM mechanisms are energetically forbidden.
In Fig. \ref{Fig_2}, zero is placed at half gap. CM lifetimes calculated using the BGF procedure extend for about $2$ eV above the CM energy threshold (we consider the same window of energies of Ref \cite{govoni_nat}). For what concern calculations performed with the PDEP methods, instead, we decided to extend such energy window to deeper values, up to about $6$ eV below the CM energy threshold.\\
\noindent
Results of Fig. \ref{Fig_2} point out that:
\begin{enumerate}
\item  CM lifetimes calculated  using the BGF and PDEP approach are very similar, 
\item CM is very fast and settle to fraction of femtosecond in the range of energies analyzed,
\item the inclusion of the screened  part of the Coulomb interaction is very important: when the second term in  Eq.  \eqref{W} is not  considered, 
 CM lifetimes are under-estimated of about one order of magnitude.
\end{enumerate}
Point one confirms the robustness and reliability of the two procedures adopted in the calculation of CM processes in Si-NCs. Depending on the size and characteristic of the system and on the features of the platform where simulation are performed (number of nodes, number of core per nodes, memory per core,  parallel architecture typology, etc..), one method can be more efficient than the other, but if the convergence is correctly achieved (truncation  of the reducible density-density $\chi_{{\bf{G}}, {\bf{G}}'}$ and number of PDEP basis), the two procedures lead to similar results.
The small differences detectable in Fig. \ref{Fig_2} (CM lifetimes are slightly lower when the PDEP procedure is adopted, both for the bare and the bare+screened cases)
are probably due to the different size of the cell used in the simulations ($90$ a.u, that is $3.66$ times the diameter of the NCs for the calculation performed with the BGF technique,  $70$ a.u, that is $2.85$ times the diameter of the NC for the calculation performed with the PDEP procedure)  and by the fact that 
the BDF methods were applied by adopting an exact box-shaped Coulomb cut-off technique  in order to remove the spurious Coulomb interaction among replicas (see Ref. \cite{rozzi}). 
\subsection{OH-terminated Si-NCS, role played by the surface passivation.}
\begin{figure}[b]%
\includegraphics*[width=\linewidth]{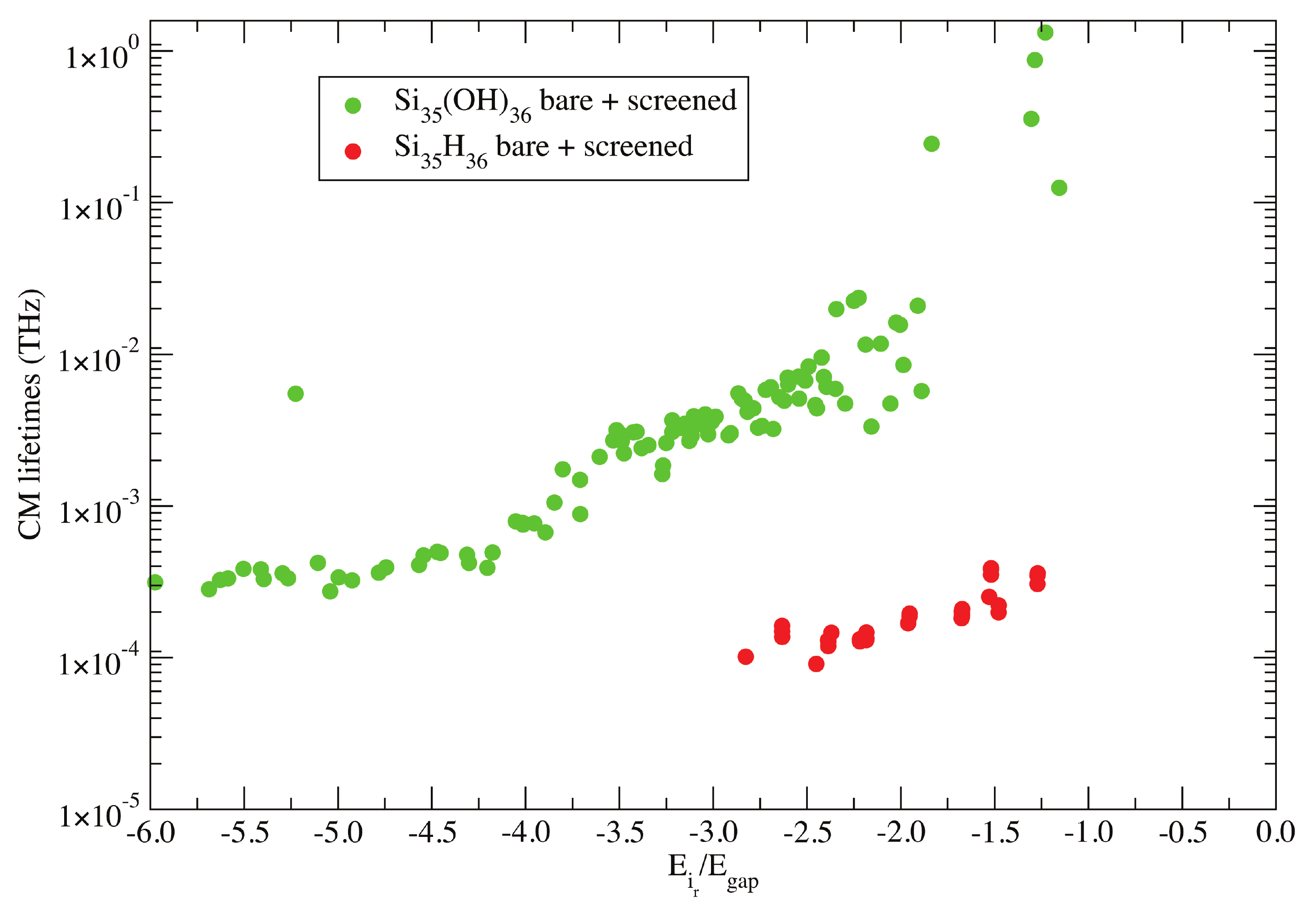}
\caption{%
CM lifetimes for both the $\text{Si}_{35}\text{(OH)}_{36}$ and the $\text{Si}_{35}\text{H}_{36}$ are reported adopted a relative energy scale.}
\label{Fig_4}
\end{figure}

In this section we extend our analysis to the study of CM processes in OH- terminated Si-NCs.  A direct comparison with the results obtained in section \ref{H_term} will permit to shed light on the role played by the passivation on the CM activity.
As pointed out by different Authors, H-passivation represents a very simple NC surface termination that is, in some cases, far from the one observed in realistic sample. For this reason we investigate a more complicated system, that is the $\text{Si}_{35}\text{(OH)}_{36}$, where the structures is obtained from H- terminated Si-NCs by replacing hydrogens with hydroxyls functional groups (-OH); the NCs is then relaxed. 
The choice of OH groups as Si-NCs passivation has several reasons. As proven in Ref. [33], beside strain effects, there are important similarities between the calculated electronic and optical properties for silica embedded Si-NCs and the Si-NCs passivated with OH groups. Moreover this passivation allows to take contact with experiments performed on all-inorganic colloidal Si-NCs terminated by OH.\\
\noindent
Following section \ref{H_term},  we consider only mechanisms ignited by hole relaxation, although for this system the energy conservation rule allows also mechanisms ignited by electron relaxation. 
\begin{figure}[t]%
\includegraphics*[width=\linewidth]{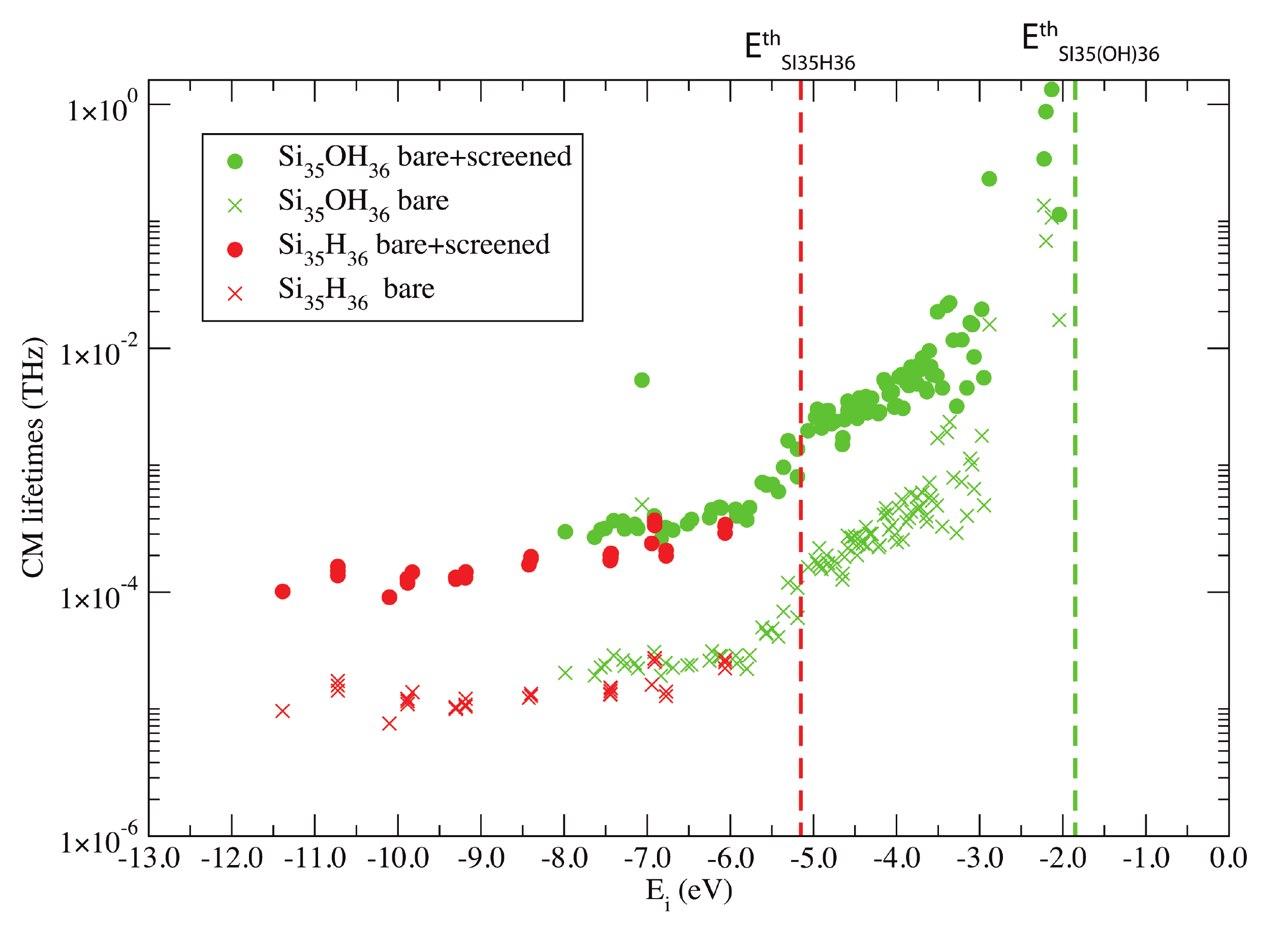}
\caption{%
CM lifetimes for both the $\text{Si}_{35}\text{(OH)}_{36}$ and the $\text{Si}_{35}\text{H}_{36}$ are reported adopted an absolute energy scale. Red and green vertical dashed lines denote the CM energy threshold for the H- and OH- terminated Si-NC, respectively.}
\label{Fig_3}
\end{figure}
CM lifetimes for the $\text{Si}_{35}\text{(OH)}_{36}$ are obtained using the PDEP procedure, with $2100$  vectors basis for the calculation of the symmetrized irreducible polarizability. Calculated CM lifetimes are reported in  Fig. \ref{Fig_3} (absolute energy scale) and in Fig. \ref{Fig_4} (relative energy scale) and compared with the one obtained for the   $\text{Si}_{35}\text{H}_{36}$. \\
\noindent
In  Fig. \ref{Fig_3} both the pure bare (crosses) and the bare+screened (dots) contributions are reported.  Due to the different energy gaps ($3.42$ eV for the $\text{Si}_{35}\text{H}_{36}$ and  $1.23$  eV for the $\text{Si}_{35}\text{(OH)}_{36}$ ), NCs show different activation CM energy threshold $E_{Si_{35}H_{36}}^{th}$ and $E_{Si_{35}(OH)_{36}}^{th}$, that are identified by the red and the green vertical dashed lines, respectively (see  Fig. \ref{Fig_3}). Obviously, CM is theoretically permitted at lower energies when the OH passivation is taken into account.\\
\noindent
Two important particulars emerge from the analysis of plots of Fig. \ref{Fig_3}. First at all, also for the $\text{Si}_{35}\text{(OH)}_{36}$, a detailed evaluation of the CM lifetimes require the calculation of the screened term $W^{p}_{{\bf{G}}, {\bf{G}}'}({\bf{q}}, 0)$  in Eq. \eqref{W}, and in particular a detailed treating of the local fiels effects that cannot be evaluated by using oversimpify  models.  By only including the bare term in the  screened interaction, we have an  under-estimation of the CM lifetimes of approximately one order of magnitude. Secondly,  CM lifetimes decrease more rapidly in the   $\text{Si}_{35}\text{H}_{36}$ with respect  the $\text{Si}_{35}\text{(OH)}_{36}$ and settle to femtoseconds just few tenths of eV below $E_{Si_{35}H_{36}}^{th}$.
Remarkably, at low energies, that is far from the activation threshold $E_{\text{Si}_{35}\text{(OH)}_{36}}^{th}$, CM lifetimes seems to be almost independent on the passivation \cite{Nota_PSS}.
From $-6$ eV to $-8$ eV, for instance, red and green dots (crosses) are almost overlapped (starting from the CM energy threshold, for both the considered NCs, we have calculated CM lifetimes in a window of energy of about $6$ eV. Due to the huge number of possible CM decay paths, number that increase when the range of considered energies $E_i$ increase, we cannot extend  the range of energies considered in the calculations, and for the ${\text{Si}_{35}\text{(OH)}_{36}}$ NC, we cannot consider CM transitions for energy $E_i$ below  $\approx -8$ eV).
Noticeably $E_i$ denotes the energy of the carrier igniting the process calculated with respect the midgap. As a consequence, when NCs of different $E_g$ are considered,  $E_i$ identifies carriers that are placed at a different distance with respect the band edge. For this motive we introduce a new parameter, called $E_{{i}_{r}}$, that represents the distance in energy between the carrier igniting the CM process and the band edge (in our specific case the valence band edge).
For mechanisms ignited by hole relaxation we have:  $ E_{{i}_{r}}= -(\mid E_{i} \mid -\mid E_{gap}\mid /2)$. For the systems considered in this work we have $\mid E_{{i}_{r}}^{Si_{35}H_{36}} \mid < \mid  E_{{i}_{r}}^{Si_{35}(OH)_{36}}\mid $. As pointed out by G. Allan and C. Delerue,  CM rates are proportional to the  product of an effective screened Coulomb matrix element and of the density of final states at the energy $E_i$ ($\rho_f(E_i)$) (for more details see Ref. \cite{delerue_impact1}).  $\rho_f(E_i)$ depends on the electronic properties of the system and increase with $E_{{i}_{r}}$. In general, $\rho_f(E_i)$ is greatest in the OH-terminated system than in the H-terminated one, that is: $\rho_f(E_i)_{Si_{35}(OH)_{36}} > \rho_f(E_i)_{Si_{35}H_{36}}$. At the same time, far from the activation threshold, for a fixed value of $E_i$, H-terminated Si-NCs show higher effective screened Coulomb matrix elements as a consequence of the strong confinement of the charge density induced by the hydrogens. 
Far for the activation threshold therefore,  we have a sort of compensation between these two effects (Coulomb interaction higher in the H-terminated system with respect the OH- terminated one and  $\rho_f(E_i)$ higher in the OH-terminated systems with respect the H- terminated one) that leads, at a fixed values of $E_i$, to similar CM lifetimes for the $\text{Si}_{35}\text{(OH)}_{36}$and the $\text{Si}_{35}\text{H}_{36}$.\\
\noindent
Results of panel (a) are not sufficient to understand the effects induced by CM on the population of the excited states.  To be effective, CM has to dominate over all the other recombination and relaxation mechanisms,  and in particular the hot carrier cooling via phonon emission. 
In  Fig. \ref{Fig_4},  we report the calculated CM lifetimes as a function of the ratio $E_{{i}_{r}}/E_{gap}$, that is by adopting the so called relative energy scale (here the sign minus indicates that processes are induced by hole relaxation). As pointed out by M. C. Beard et al. Ref. \cite{beard_ncvsbulk}, this plot provide more information about the competition between CM with other energy relaxation channels. When this scale is adopted, we observe than CM is more efficient in 
$\text{Si}_{35}\text{H}_{36}$  than in $\text{Si}_{35}\text{(OH)}_{36}$. The presence of oxygen at the surface, therefore, can reduce the relevance of the CM processes.
\section{Conclusions}
In this work we have calculated CM lifetimes in Si-NCs. Our calculations have been performed using two different (largely diffused) methodologies, which permit to calculate the screened coulomb interaction by using the non-interacting electron-hole Green's function method and the projective eigendecomposition of the dielectric screening procedure. For the considered systems, the two methodologies lead to very similar results. Both of them point out the importance of a detailed description of the screening caused by the medium and of the local fields terms, that are fundamental when the  screened Coulomb interaction is calculated. We have then compared CM lifetimes of H-terminated and OH-terminated  Si-NCs. We have proven that the presence of oxygen at the surface reduce the theoretical CM activation threshold. When an absolute energy scale is adopted, far from the activation threshold, CM lifetimes seems to be independent on the passivation. In order to have a more general comprehension of the effect,  we have also analyzed CM decay path adopting a relative energy scale, that permit to have more information about the competition between CM with other energy relaxation channels. When a relative energy scale is adopted we observe that CM processes are more efficient in H-terminated than in OH-terminated Si-NCs. This work represent a first, important, step toward the comprehension of the mechanisms that connect CM decay processes with NCs passivation and open new prospective for future, more complicated, analysis of the CM activity in NCs.
\section{Acknowledgement}
The authors thank the Super-Computing Interuniversity Consortium CINECA for support and high-performance computing resources under the Italian Super-Computing Resource Allocation (ISCRA) initiative, PRACE for awarding us access to resource FERMI IBM BGQ, and MARCONI HPC cluser based in Italy at CINECA.
Ivan Marri acknowledges support/funding from European Union H2020-EINFRA-2015-1 programme under grant agreement No. 676598 project "MaX - materials at the exascale" .

\bibliographystyle{pss}
\bibliography{manuscript}

%

\end{document}